\documentclass[lineno]{jfm}

\usepackage{graphicx}
\usepackage{amsmath}
\usepackage{amssymb}
\usepackage{pifont}
\usepackage{txfonts}
\usepackage{psfrag}
\usepackage[round]{natbib}
\usepackage[T1]{fontenc}
\usepackage{lmodern}
\usepackage{tikz}
\usetikzlibrary{shapes.geometric,shapes.symbols}
\usepackage{scalerel}
\usepackage{upgreek}
\usepackage{booktabs}
\usepackage{verbatim}
\usepackage{framed}
\usepackage{fancyvrb}
\usepackage{varwidth}
\usepackage{bbding}
\usepackage{subfigure}
\usepackage{multirow}
\usepackage[normalem]{ulem}
\usepackage{multicol}
\usepackage{xcolor}
\usepackage{breqn}
\usepackage{soul}
\usepackage{dirtytalk}
\usepackage{upquote}
\usepackage{authblk}
\usepackage{array}
\newcolumntype{P}[1]{>{\centering\arraybackslash}p{#1}}

\usepackage{hyperref}
\hypersetup{
    colorlinks = true,
    urlcolor   = blue,
    citecolor  = black,
}

\newcommand{\RomanNumeralCaps}[1]

\title{On the relationship between manipulated inter-scale phase and energy-efficient turbulent drag reduction}
\author{Rahul Deshpande\aff{1}
  \corresp{\email{raadeshpande@gmail.com}},
  Dileep Chandran\aff{1},
	Alexander J. Smits\aff{2},
 and Ivan Marusic\aff{1}}

\affiliation{\aff{1}Department of Mechanical Engineering, University of Melbourne, Parkville, VIC 3010, Australia
\aff{2}Department of Mechanical and Aerospace Engineering, Princeton University, Princeton, NJ 08544, USA}

\begin{document}
\maketitle

\begin{abstract}
We investigate the role of inter-scale interactions in the high-Reynolds number skin-friction drag reduction strategy reported by \citeauthor{marusic2021}  (\textit{Nat.\ Commun.}, vol.\ 12, 2021).
The strategy involves imposing relatively low-frequency streamwise travelling waves of spanwise velocity at the wall to actuate the drag generating outer-scales. 
This approach has proven to be more energy-efficient than the conventional method of directly targeting the drag producing inner-scales, which typically requires actuation at higher frequencies.   
Notably, it is observed that actuating the outer-scales at low frequencies leads to a substantial attenuation of the major drag producing inner-scales, 
suggesting that the actuations affect the non-linear inner-outer coupling inherently existing in wall-bounded flows. 
In the present study, we find that increased drag reduction, through imposition of spanwise wall oscillations, is always associated with an increased coupling between the inner and outer scales. 
This enhanced coupling emerges through manipulation of the phase relationships between these triadically linked scales, with the actuation forcing the entire range of energy-containing scales, from the inner (viscous) to the outer (inertial) scales, to be more in-phase. 
We also find that a similar enhancement of this non-linear coupling, via manipulation of the inter-scale phase relationships, occurs with increasing Reynolds number for canonical turbulent boundary layers.
This indicates improved efficacy of the energy-efficient drag reduction strategy at very high Reynolds numbers, where the energised outer-scales are known to more strongly superimpose and modulate the inner-scales. 
Leveraging the inter-scale interactions, therefore, offers a plausible mechanism for achieving energy-efficient drag reduction at high Reynolds numbers.  
\end{abstract}


\section{Introduction}
\label{intro}

A characteristic feature of any turbulent flow is the broad range of scales, or `eddies', which carry the total kinetic energy.  The eddies are coupled with each other non-linearly, and this non-linear `cross-talk' across the energy spectrum is the cornerstone for the energy transfer process in a turbulent flow.
Since the non-linearity in the governing Navier-Stokes equations is quadratic, this inter-scale coupling is, by definition, triadic in nature \citep{duvvuri2015}.
Here, triadically coupled eddies correspond to the scenario when the time scales ($T_{i}$) of any three eddies (say $l$, $m$ and $n$) are related via either of the four relationships:
\begin{equation}
\label{eq0}
\begin{aligned}
(i){\;}{\omega_l=\omega_m-\omega_n}, \
(ii){\;}{\omega_l=\omega_m+\omega_n}, \
(iii){\;}{\omega_l=2\omega_m}, \
(iv){\;}{\omega_l=2\omega_n},
\end{aligned}
\end{equation}
where the frequencies $\omega_i=2\pi /T_i$.

\begin{figure}
   \captionsetup{width=1.0\linewidth}
  \centerline{\includegraphics[width=1.0\textwidth]{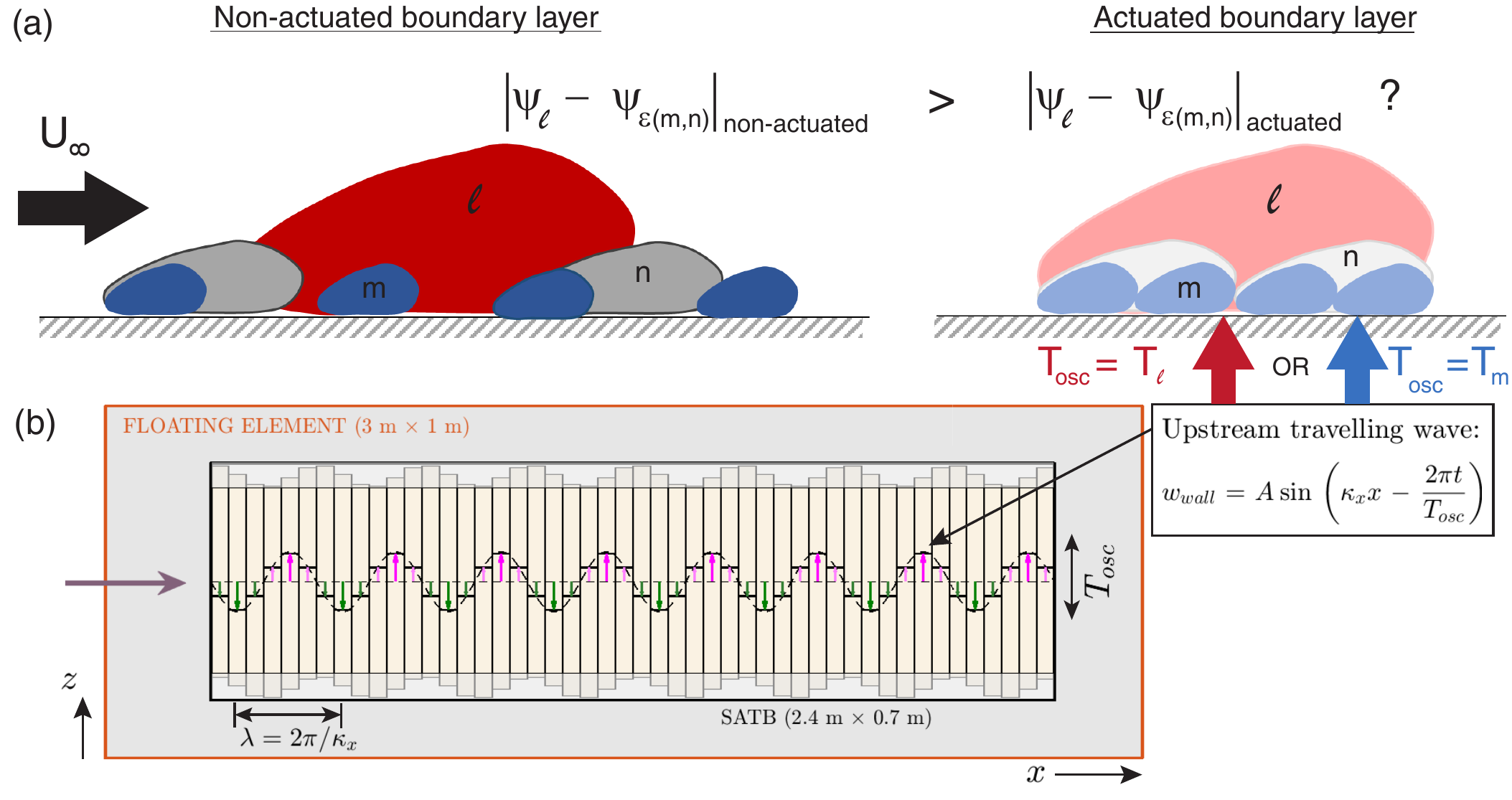}}
  \caption{(a) Schematic representation of triadically coupled eddies ($l$, $m$, $n$) with time scales $T_{l}$, $T_{m}$, $T_{n}$ and phase ${\psi}_{l}$, ${\psi}_{m}$, ${\psi}_{n}$, respectively.
	(b) Schematic of the spanwise wall-actuation scheme investigated by \citet{marusic2021} and others. The terminology is discussed in $\S$\ref{intro}.} 
\label{fig1}
\end{figure}

In the case of wall-bounded turbulent flows, the nature of inter-scale coupling has been verified previously through a set of experiments and analysis by McKeon and co-workers \citep{jacobi2013,jacobi2017,duvvuri2015,duvvuri2017}.
The experiments involved artificial excitation of a large \emph{outer-scaled} eddy, which spanned the boundary layer (consider for instance, eddy $l$ in figure \ref{fig1}), by imposition of an oscillatory forcing via a dynamic roughness element.
Statistical analysis of the velocity signals acquired within this perturbed boundary layer confirmed the influence of this forcing (with time scale, $T_{l}$) on the corresponding small \emph{inner-scaled} eddies $m$ and $n$ (figure \ref{fig1}a), which are triadically coupled with $l$.
This coupling, or `cross-talk', was found to be facilitated by a phase locking of the synthetic scale (${\psi}_{l}$) and the average triadic `envelope' of the small scales (${\psi}_{{\epsilon}(m,n)}$), where ${\psi}_{i}$ and ${\epsilon}$ respectively denote the phase and the envelope \citep{mathis2009}.

As discussed by  \cite{duvvuri2015} and \cite{jacobi2017}, this inter-scale coupling offers an  opportunity to control and manipulate the near-wall inner-scales \citep{jimenez1991,waleffe1993,hamilton1995,jimenez1999} by exciting the triadically coupled outer-scales, with potential implications for drag reduction. 
For instance, the inner-scaled eddies near the wall are a major contributor to the total skin-friction drag across the practically-relevant Reynolds number range \citep{deck2014,chandran2022}. 
It is challenging, however, to actuate these eddies owing to their very small time/length scales in physical units \citep{quadrio2011,ricco2021}.
Alternatively, it may be possible to leverage the coupling between the turbulent scales to indirectly attenuate these drag contributing inner-scales through their triadically coupled outer-scales. 

In the present study, we test this idea by considering the flow-control approach of imposing spanwise wall-oscillations \citep{akhavan1993,baron1995,choi1998,karniadakis2003,quadrio2009,agostini2014,ricco2021}, wherein the oscillating wall elements are synchronized to generate an upstream traveling wave with respect to the mean flow direction.
Figure \ref{fig1}(b) schematically describes this control strategy, where the instantaneous spanwise velocity ($w_{wall}$) imposed on the wall is given by: ${w_{wall}}(x,t)$ $=$ ${A}\:{\sin}({{\kappa}_{x}}x$ $-$ $\omega_{osc} t)$.
Here, $T_{osc}=2 \pi/\omega_{osc}$ and $A$ are the time period and amplitude of the spanwise oscillation, respectively;  ${\kappa}_{x}$ $=$ 2$\pi$/${\lambda}$ is the streamwise wavenumber of the traveling wave where $\lambda$ is the wavelength; $t$ denotes time; and 
$u$, $v$ and $w$ denote the velocity fluctuations along the streamwise ($x$), wall-normal ($y$) and spanwise ($z$) directions, respectively.
This approach of imposing spanwise oscillations on the wall has been investigated extensively (refer to \citet{ricco2021} and references therein), predominantly for its ability to achieve significant drag reduction (DR) through actuating the near-wall inner-scaled motions having a characteristic viscous-scaled time scale of $T^+ \approx 100$, which we refer to here as the \emph{inner-scaled actuation} (ISA) strategy \citep{rouhi2022}.
These investigations have mostly been limited to low Reynolds number flows, for which the inner-scales are the dominant (if not sole) contributors to the total drag \citep{schoppa2002,kim2011}.
Oscillating these wall elements at the associated small time scales ($T_{osc}^+ \lesssim 100$), however, incurs a large power cost (${\propto}\,({T_{osc}})^{-\frac{5}{2}}$; \citealp{quadrio2011}), making net power savings from the ISA strategy less likely \citep{rouhi2022}.
Considering the inverse relationship between power cost and oscillation time period, previous studies performed at low Reynolds numbers (\citealp{gatti2016}; some cases in \citealp{marusic2021})  have attempted to reduce skin-friction drag by targeting the outer-scales, associated with large $T^{+}_{osc}$.
However, these attempts were proven less effective at such low Reynolds numbers, wherein the contributions of outer-scales to the skin-friction drag are statistically insignificant \citep{chandran2022}.
Based on the premise that outer-scale contributions to the turbulent skin-friction increase with Reynolds number, \citet{marusic2021} demonstrated that large-time scale spanwise wall-actuation targeting these outer-scales can, however, yield DR for a sufficiently high Reynolds number flow.
More importantly, this \emph{outer-scaled actuation} (OSA) strategy  \citep{chandran2022} required considerably lower input power, thereby yielding drag reduction with net power savings.
The exact mechanism behind this new drag reduction strategy, however, still remains unknown.

The present study investigates whether this energy-efficient drag reduction via the OSA strategy is facilitated by the manipulation of the inter-scale phase of the triadically coupled inner- and outer-scales.
To this end, statistical analysis, based on the arguments of McKeon and co-workers \citep{duvvuri2015,jacobi2017}, is conducted on the experimental data of \citet{chandran2022}. 
The same analysis is also extended for the ISA strategy to confirm whether the triadically coupled scales can be influenced from either end of the frequency spectrum.
Following \citet{marusic2021}, we consider oscillation periods, $T^{+}_{osc}$ $\lesssim$ 350 to be associated with ISA strategy, while $T^{+}_{osc}$ $\gtrsim$ 350 with the OSA strategy, where $T^{+}_{c}$ $=$ 350 is taken to be the nominal demarcation between inner and outer scales.

\section{Experimental data}
\label{exp}

All experiments reported in this study were performed in the large wind tunnel (HRNBLWT) at the University of Melbourne. 
It features a 3 m$\times$1$\:$m (length $\times$ width) floating element balance (figure \ref{fig1}b)  located at $19.5\:\mathrm{m} \lesssim x \lesssim 22.5 \:\mathrm{m}$ downstream of the start of the test section.
Due to this long development length, the boundary layer properties, and therefore the Reynolds number, remain almost constant along the 3 m length of the floating element \citep{talluru2013}.
The surface actuation test bed (SATB) is flush mounted in a rectangular cutout of 2.7$\:{\rm m}{\:}\times{\:}0.7{\:}$m (length $\times$ width) centred within this floating element.
The SATB is a combination of 4 independently-controlled, servo motor driven machines that run in a phase-synchronised manner to impose a 8$\lambda$ long upstream travelling sinusoidal wave at the wall (figure \ref{fig1}b).
This is made possible by discretising two sinusoidal waves into 12 slats (oscillating elements) connected to a single machine, which cumulatively span a streamwise length of 2$\lambda$ = 0.6$\:$m and have their phase controlled by a central camshaft.
These slats, along with the camshaft, were fabricated with a high degree of precision and tolerance, enabling a spanwise wall motion while maintaining a negligible gap ($\sim$100${\:}{\mu}$m) between the individual slats.
For non-actuated cases, a hydraulically smooth boundary layer flow was ensured \citep{marusic2021} and the data were found to match with previously published smooth-wall data from the Melbourne wind tunnel at the same location \citep{talluru2013,marusic2015}.
When actuated, the slats have a fixed half stroke length of $d$ = 18$\:$mm and can be oscillated at frequencies as high as $1/{T_{osc}}$ = 25$\:$Hz, leading to a maximum achievable spanwise velocity of $A$ (= $2{\pi}d/{T_{osc}}$) = 2.83 m/s.
It is important to note here that a discretized imposition of the streamwise travelling sinusoidal waves inevitably introduces high wavenumber harmonics  of low amplitudes on the turbulent boundary layer \citep{auteri2010}. 
However, considering that the present experiments regard fundamental modes with very low wavenumbers (${{\kappa}^{+}_{x}}$ $\lesssim$ 0.0014; table \ref{tab1}), the corresponding higher harmonics are expected to only marginally influence the drag reduction estimates \citep{gatti2016,chandran2022}.
Further details regarding the characterization and validation of the SATB can be found in \citet{marusic2021} and \citet{chandran2022}.

\begin{table}
   \captionsetup{width=1.0\linewidth}
\centering
\begin{center}
\begin{tabular}{@{\extracolsep{1pt}}P{1.1cm} P{0.6cm} P{0.6cm} P{0.4cm} P{0.8cm} P{0.9cm} P{0.9cm} P{1.5cm} P{0.9cm} P{1.2cm} P{0.5cm} P{1cm} }
\multicolumn{4}{c}{TBL at hotwire} & \multicolumn{6}{c}{Parameters associated with spanwise} & \multicolumn{1}{c}{Drag} & \multicolumn{1}{c}{Net power} \\
\multicolumn{4}{c}{measurement location} & \multicolumn{6}{c}{oscillations imposed on the wall} & \multicolumn{1}{c}{reduction} & \multicolumn{1}{c}{savings} \\
\cmidrule{1-4} \cmidrule{5-10} \cmidrule{11-12}
$Re_{{\tau}}$ & $U_{\infty}$ & $\delta$ & $x$ & $1/{T_{osc}}$ & $A$ & ${\kappa}_{x}$ & $T^{+}_{osc}$ & $A^+$ & ${\kappa}^{+}_{x}$ & DR(\%) & NPS(\%)\\
 & (m/s) & (m) & (m) & (Hz) & (m/s) & (1/m) & $=\frac{{T_{osc}}\,{U^{2}_{{\tau}_{o}}}}{\nu}$ & $=\frac{A}{{U_{{\tau}_{o}}}}$ & $=\frac{{{\kappa}_{x}}{\nu}}{{U_{{\tau}_{o}}}}$ &  &  \\
\hline
\underline{\textbf{Non-actuated (i.e. canonical) flow:}} & & & & & & & & & &\\
2 800$^*$ & 20$^*$ & 0.06$^*$ & 1.60$^*$ & $-$ & $-$ & $-$ & $-$ & $-$ & $-$ & $-$ & $-$\\
13 000$^*$ & 20$^*$ & 0.30$^*$ & 17.5$^*$ & $-$ & $-$ & $-$ & $-$ & $-$ & $-$ & $-$ & $-$\\
6 000 & 7 & 0.39 & 21 & $-$ & $-$ & $-$ & $-$ & $-$ & $-$ & $-$ & $-$\\
9 700 & 11 & 0.39 & 21 & $-$ & $-$ & $-$ & $-$ & $-$ & $-$ & $-$ & $-$\\
\underline{\textbf{Inner-scaled actuation (ISA; $T^+$ $\lesssim$ 350):}} & & & & & & & & & & \\
6 000 & 7 & 0.39 & 21 & 10 & 1.13 & 20.94 & 348 & 4.9 & 0.0014 & 10\% & +4.8\%\\
6 000 & 7 & 0.39 & 21 & 15 & 1.70 & 20.94 & 232 & 7.4 & 0.0014 & 16\% & +1.6\%\\
6 000 & 7 & 0.39 & 21 & 20 & 2.26 & 20.94 & 174 & 9.8 & 0.0014 & 20\% & -5.9\%\\
6 000 & 7 & 0.39 & 21 & 25 & 2.83 & 20.94 & 140 & 12.3 & 0.0014 & 24\% & -21.5\%\\
\underline{\textbf{Outer-scaled actuation (OSA; $T^+$ $>$ 350):}} & & & & & & & & & & \\
9 700 & 11 & 0.39 & 21 & 10 & 1.13 & 20.94 & 906 & 3.0 & 0.0008 & 9.5\% & +8.3\%\\
9 700 & 11 & 0.39 & 21 & 15 & 1.70 & 20.94 & 604 & 4.6 & 0.0008 & 11.5\% & +8.4\%\\
9 700 & 11 & 0.39 & 21 & 20 & 2.26 & 20.94 & 453 & 6.2 & 0.0008 & 12.5\% & +6.2\%\\
9 700 & 11 & 0.39 & 21 & 25 & 2.83 & 20.94 & 362 & 7.8 & 0.0008 & 15\% & +3.8\%\\
\hline
\end{tabular}
\caption{A summary of the experimental dataset employed in this study, originally reported in \citet{marusic2021} and \citet{chandran2022}.
Parameters with $^*$ are associated with the published hotwire dataset of \citet{marusic2015}.
The turbulent boundary layer (TBL) characteristics reported in the first three columns are for a non-actuated scenario.}
\label{tab1}
\end{center}
\end{table}

All velocity data analyzed in this study were acquired using hot-wire anemometry for various non-actuated and actuated cases (140 $\lesssim$ $T^{+}_{osc}$ $\lesssim$ 906; table \ref{tab1}). 
Hot-wire data were acquired with an actuation length of $\gtrsim$ 2${\lambda}$ ($\gtrsim$ 2$\delta$), beyond which the local drag modified due to the actuation was found to have nearly saturated \citep{chandran2022}, consistent with past observations in the literature \citep{ricco2004,skote2019wall}. 
The mean wall shear stress for the non-actuated ($\overline{\tau_{w_o}}$) and actuated flow cases ($\overline{\tau_{w}}$) was measured directly using the floating element drag balance, and also estimated from dedicated hot-wire experiments conducted in the viscous sublayer, to estimate DR = (1 - ${\overline{\tau_{w}}}$/$\overline{\tau_{w_o}}$). 
Table \ref{tab1} reports the DR\% as well as the percentage net power savings (NPS\%) associated with each actuated case, where the latter is essentially the difference between DR\% and the percent net input power required to move the flow sideways.
Here, NPS was computed based on the generalised Stokes layer theory (GSL; \citealp{quadrio2011}), which estimates the net input power required by an `ideal' actuation system (i.e. neglecting any mechanical losses) to generate a Stokes layer.
Interested readers may refer to appendix A, where we have provided further details regarding estimation of NPS.
As expected, NPS was found to decrease with decreasing $T^+_{osc}$, with positive NPS accompanied by significant DR predominantly for the OSA cases \citep{chandran2022}.
Detailed description of the present measurements can be found in \citet{marusic2021}, \citet{rouhi2022} and \citet{chandran2022}, where the data are analysed in greater depth.

The present analysis is limited to boundary layers at $Re_{{\tau}}$ ($={{\delta}{U_{{\tau}_{o}}}/{\nu}}$) $\approx$ 6000 and 9700. 
Here, $\rho$ and $\nu$ are the density and kinematic viscosity of air, respectively, while ${\delta}$ and ${U_{{\tau}_{o}}}=\sqrt{\overline \tau_{w_o}/\rho}$ are the boundary layer thickness and friction velocity associated with the non-actuated flow.
The superscript `+' will denote normalization in viscous units using $\nu$ and $U_{{\tau}_{o}}$. 
We investigate the inter-scale phase relationships for various ISA and OSA cases at these two high Reynolds numbers by analyzing streamwise velocity fluctuations ($u$), acquired by a normal hot-wire positioned at 4.5 $\lesssim$ $y^+$ $\lesssim$ 6 (where $y^+$ = $y{U_{{\tau}_{o}}}/{\nu}$).
This $y^+$-range within the linear region was deemed optimal in a way that is close enough to the wall, for the $u$-signal to be sensitive to spanwise wall oscillations, while also being far enough to have a substantial streamwise variance, $\overline{u^2}$, for a greater signal-to-noise ratio (overbar denotes time averaging). 
Additionally, in this $y^+$-range, any undesired wall-conduction effects due to the close proximity of the hot-wire probe to the wall were found to be minimal.
In the forthcoming section, we present statistical comparisons between actuated and non-actuated cases (for a given $Re_{\tau}$) at matched $y^+$, to bring out the `absolute' response of the flow to wall oscillation \citep{agostini2014}. 
However, the same trends are noted for statistical comparisons at matched $y^*$ $=$ $y{U_{\tau}}/{\nu}$, i.e. based on the local friction velocity $U_{\tau}$.

\section{Results}
\label{results}

\subsection{Mean phase between all triadically coupled scales}

\citet{duvvuri2015} showed that the phase relationships between triadically coupled scales (coexisting at any $y$) can be estimated by computing the skewness of the streamwise velocity fluctuations (${\mathcal{S}}_{u}$) at $y$.
This was demonstrated by decomposing the experimentally acquired, statistically stationary $u$-time series ($u$($t$)) as a summation of its Fourier modes, i.e. 
$u(t)$ $=$ ${\sum_{i = 1}^{\infty}}{{{\alpha}_{i}}\sin({{{\omega}_{i}}t} + {{\psi}_{i}})}$, with circular frequencies ${\omega}_{i}$ ($=$ 2${\pi}$/${T_{i}}$), amplitudes (${\alpha}_{i}$), phase (${\psi}_{i}$) and 0 $<$ ${\omega}_{i}$ $<$ ${\omega}_{\infty}$.
That is, 
\begin{equation}
\label{eq2}
\begin{aligned}
{{\mathcal{S}}_{u}} = \frac{\overline{u^3}}{{\sigma}^3} = {{\frac{6}{4{{\sigma}^3}}}\sum_{\substack{ \forall \, l,m,n\\{{\omega}_{l}} < {{\omega}_{m}} < {{\omega}_{n}}\\{{\omega}_{l}} + {{\omega}_{m}} = {{\omega}_{n}} }} {{\alpha}_{l}}{{\alpha}_{m}}{{\alpha}_{n}}\;{\sin({{\psi}_{l}} + {{\psi}_{m}} - {{\psi}_{n}})}} + {{\frac{3}{4{{\sigma}^3}}}\sum_{\substack{ l=1\\{{\omega}_{n}} = 2 {{\omega}_{l}} }} {{\alpha}^2_{l}}{{\alpha}_{n}}\;{\sin(2{{\psi}_{l}} - {{\psi}_{n}})}},
\end{aligned}
\end{equation}
where $\sigma$ $=$ $\sqrt{ \overline{u^2} }$, and ${\psi}_{l}$ + ${\psi}_{m}$ -- ${\psi}_{n}$ represents the phase difference between the triadically consistent scales, existing across the energy spectrum: $0$ $<$ ${\omega}_{i}$ $<$ ${\omega}_{\infty}$.
Hence, ${\mathcal{S}}_{u}$ can be considered as a surrogate of the average measure for the phase between the various turbulent eddies/scales coexisting at $y$.

\begin{figure}
   \captionsetup{width=1.0\linewidth}
  \centerline{\includegraphics[width=0.85\textwidth]{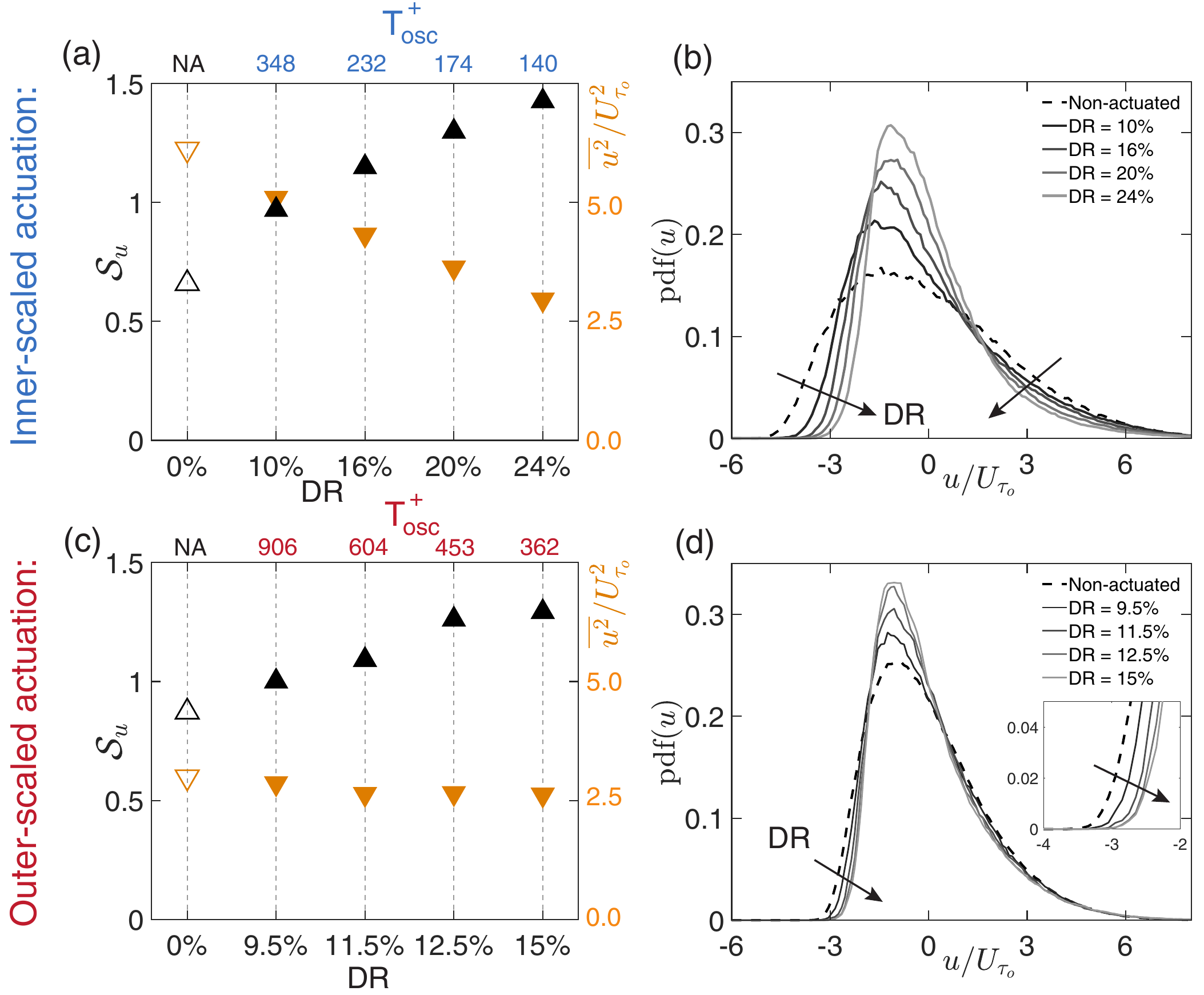}}
  \caption{(a,c) Variance ($\overline{u^2}$), skewness ($\mathcal{S}_{u}$) and (b,d) probability distribution function of the $u$-fluctuations corresponding to various (a,b) ISA ($y^+$ $\sim$ 6) and (c,d) OSA cases ($y^+$ $\sim$ 4.5).
	In (a,c), DR = 0\% or $T^{+}_{osc}$ = NA refers to the non-actuated case (empty symbols).
	$\overline{u^2}$ in (a,c) and $u$-fluctuations in (b,d) are normalized by $U_{{\tau}_{o}}$.
	Note that the abscissa is not linear in (a,c).}
\label{fig2}
\end{figure}

Figures \ref{fig2}(a,c) display our results for the near-wall values of ${\overline{u^2}}$ and ${\mathcal{S}}_{u}$. 
As expected, in scenarios of drag reduction, near-wall $\overline{u^2}$ reduces with increases in DR for both ISA and OSA cases \citep{chandran2022}.
The degree of reduction, however, is more gradual for OSA cases than for ISA cases, and it appears to be directly related to the amount of drag reduction. 
We note that the OSA cases considered here have relatively lower $A^+$ levels when compared to the ISA cases (table \ref{tab1});
hence, the lower `effectiveness' of the OSA approach is a manifestation of the parameter
space investigated and shouldn’t be associated with the overall success of the strategy.
The ${\overline{u^2}}$ trend confirms that $u$ near the wall is indeed influenced by wall actuation and hence, can be used to understand the change in flow physics with DR.

We also see from figures \ref{fig2}(a,c) that ${\mathcal{S}}_{u}$ increases with DR for both ISA and OSA cases, indicating that an increase in DR is associated with an average decrease in phase between the triadically consistent scales (see also \citealp{duvvuri2015}). 
A similar increase in ${\mathcal{S}}_{u}$ with DR has been noted previously for a drag reduced flow obtained on imposition of spanwise wall oscillations \citep{baron1995,choi1998}, or introduction of microbubbles and polymers in the flow \citep{pal1989}, but the association with inter-scale phase relationships has not been discussed.
The probability distribution functions ($pdf$) for the various $u$-signals, shown in figures \ref{fig2}(b,d), indicate that this DR trend is coupled with the reduction in intense $-u$ events ($u/{U_{{\tau}_{o}}}$ $\lesssim$ --3). 
For the ISA cases this can be associated with weakening of the low-speed near-wall streaks, as noted in some earlier works \citep{akhavan1993,agostini2014,ricco2021}, but that a similar result  can be observed for the OSA cases has only been noted recently by \citet{chandran2022}, based on near-wall PIV measurements.

\subsection{Mean phase between triadically coupled inner and outer scales}

Whether this reduction in intense $-u$ events, for the cases of OSA, exists due to the attenuation of viscosity-dominated near-wall (inner) scales, or that of the inertia-dominated outer eddies, can be understood by computing the individual inner ($u_{i}$) and outer ($u_{o}$) contributions to the $u$-fluctuations. 
In particular, we can investigate the change in mean phase between the outer ($u_{o}$ $=$ $u$($T^+$ $\gtrsim$ $T^+_{c}$)) and the inner scales ($u_{i}$ $=$ $u$($T^+$ $\lesssim$ $T^+_{c}$)), where $T^+_{c}$ = $\frac{2{\pi}}{{\omega}^{+}_{c}}$ = 350.
For this, we follow \citet{mathis2011} and linearly decompose $u$ $=$ $u_{o}$ $+$ $u_{i}$ to obtain 
\begin{equation}
\label{eq3}
\begin{aligned}
{{\mathcal{S}}_{u}} = \frac{\overline{u^3}}{{\sigma}^3} = 
 \overline{\overline{ {u}^3_{o} }} + \overline{\overline{ {u}^3_{i} }} + 3{\overline{\overline{ {{u}^2_{o}}{{u}_{i}} }}} + 3{\overline{\overline{ {{u}^2_{i}}{{u}_{o}} }}},
\end{aligned}
\end{equation}
where the double overbar denotes a time averaged quantity normalized by ${\sigma}^3$.
\citet{duvvuri2015} reported exact expressions for the four individual terms given in (\ref{eq3}), where
\begin{equation}
\label{eq4}
\begin{aligned}
{\overline{{\overline{ {u^2_{i}}{u_{o}} }}}} = {\frac{1}{2{{\sigma}^3}}}\sum_{\substack{ \forall \, l,m,n\\{{\omega}_{n}} - {{\omega}_{m}} = {{\omega}_{l}}\\0 < {{\omega}_{l}} < {{\omega}_{c}}\\{{\omega}_{m}}, {{\omega}_{n}} > {{\omega}_{c}} }} {{\alpha}_{l}}{{\alpha}_{m}}{{\alpha}_{n}}\;{\sin({{\psi}_{l}} + {{\psi}_{m}} - {{\psi}_{n}})},
\end{aligned}
\end{equation}
meaning that the cross-term $3{\overline{\overline{ {u^2_{i}}{u_{o}} }}}$ represents the mean phase difference between the outer-scale ${\omega}_{l}$ and the `envelope' of the triadically coupled inner-scales ${{\omega}_{n}}$, ${{\omega}_{m}}$.
Enhanced values for this cross-term are therefore associated with a reduction in phase between the outer and inner scales. In the remainder of this manuscript, we will therefore interpret the increase in 3${\overline{\overline{ {{u}^2_{i}}{{u}_{o}} }}}$ as an enhancement of inter-scale communication.
The other cross-term in \eqref{eq3}, $3{\overline{\overline{ {u^2_{o}}{u_{i}} }}}$, also theoretically represents an inner-outer coupling in some form, but \citet{mathis2011} found that its contributions to ${{\mathcal{S}}_{u}}$ were negligible for a canonical boundary layer.

\begin{figure}
   \captionsetup{width=1.0\linewidth}
  \centerline{\includegraphics[width=1.0\textwidth]{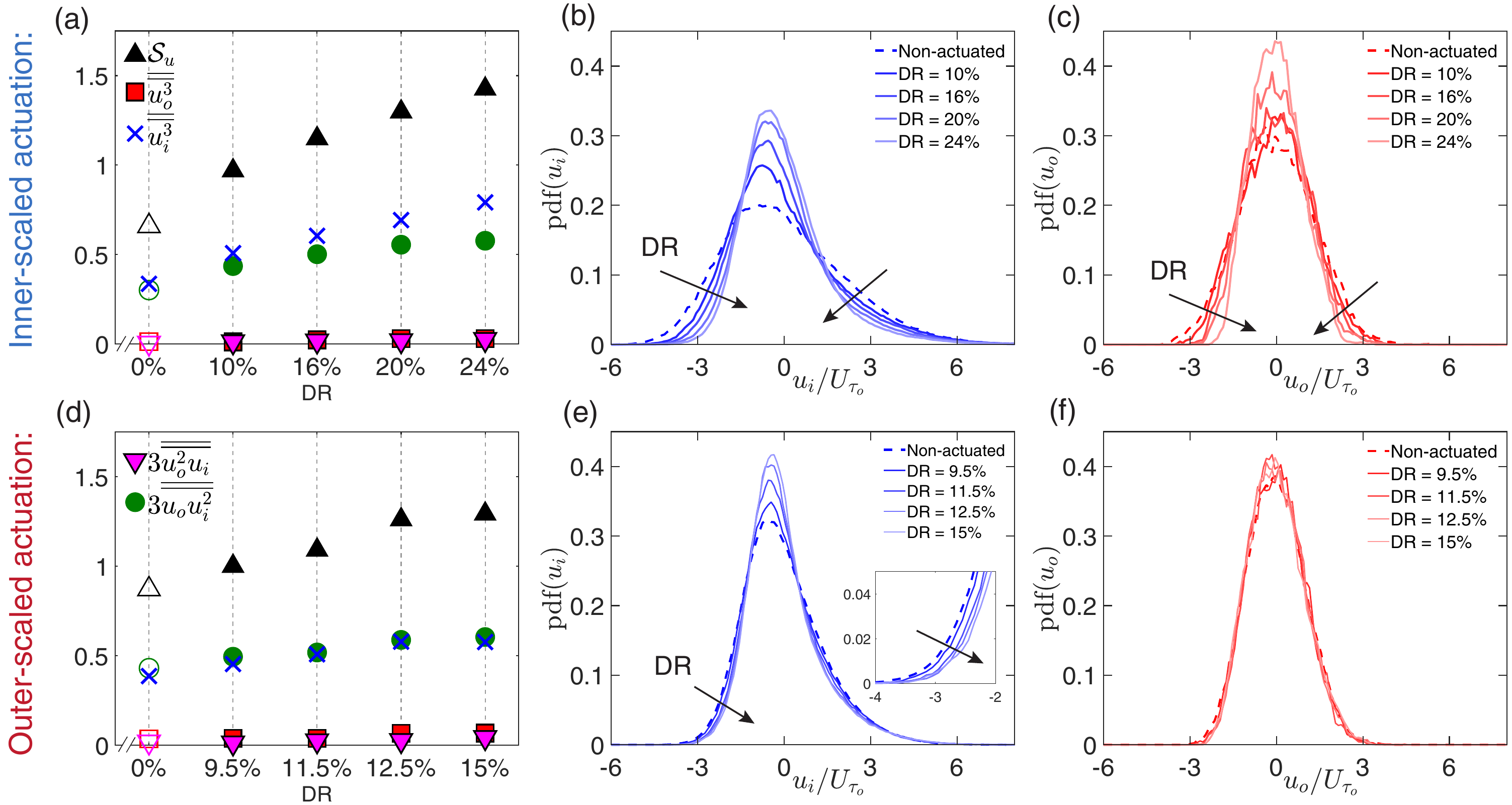}}
  \caption{(a,d) Terms obtained from decomposition of skewness ($\mathcal{S}_{u}$) corresponding to various (a) ISA and (d) OSA cases documented in table \ref{tab1}, noting that the abscissa is not linear.
	Probability distribution functions of the (b,e) inner-scale and (c,f) outer-scale sub-components of $u$, corresponding to the same (b,c) ISA and (e,f) OSA cases as in (a,d).
	In (a,d), DR = 0\% refers to the non-actuated case.
		}
\label{fig3}
\end{figure}

The four decomposed terms of ${\mathcal{S}}_{u}$ \eqref{eq3} for the various cases are shown in figures \ref{fig3}(a) and \ref{fig3}(d).
We conclude that the increase in ${\mathcal{S}}_{u}$ with DR can be attributed predominantly to two terms:
(i) $\overline{\overline{u^3_{i}}}$, which represents the average phase between the triadically coupled inner scales ($T^+$ $<$ 350), and 
(ii) 3$\overline{\overline{{u^2_{i}}{u_{o}}}}$, the cross-term that represents the average phase between the outer scales and the envelope of the triadically coupled inner-scales (\ref{eq4}).
An increase in 3$\overline{\overline{{u^2_{i}}{u_{o}}}}$ with DR suggests that drag reduction is associated with a decrease in the average phase difference between $u_{o}$ and $u_{i}$ \citep{duvvuri2015}.
In the case of ISA, the $pdf$ plots for the $u_{i}$ and $u_{o}$ signals (figures \ref{fig3}b,c) show a reduction in intense $-{u_{i}}$ events with an increase in DR, which suggests the weakening of near-wall low-speed streaks. 
But this process is also accompanied by a reduction in intense $-{u_{o}}$ and $+{u_{o}}$ events, indicating an enhanced inter-scale communication between the inner and outer scales.
Similarly, in the case of OSA (figures \ref{fig3}(e,f)), the intense $-{u_{i}}$ events are attenuated more significantly than the corresponding ${u_{o}}$ events.
The substantial drag reduction noted in the cases of both ISA and OSA strategies, thus, is a consequence of the attenuation of both -- the inner and outer scales -- which contribute to the total drag.
This is investigated further in the next section by examining the changes to the scale-specific energy (i.e. the $u$-spectra) due to spanwise wall-actuation.

\subsection{Scale-specific phase between inner and outer scales}

\begin{figure}
   \captionsetup{width=1.0\linewidth}
  \centerline{\includegraphics[width=1.0\textwidth]{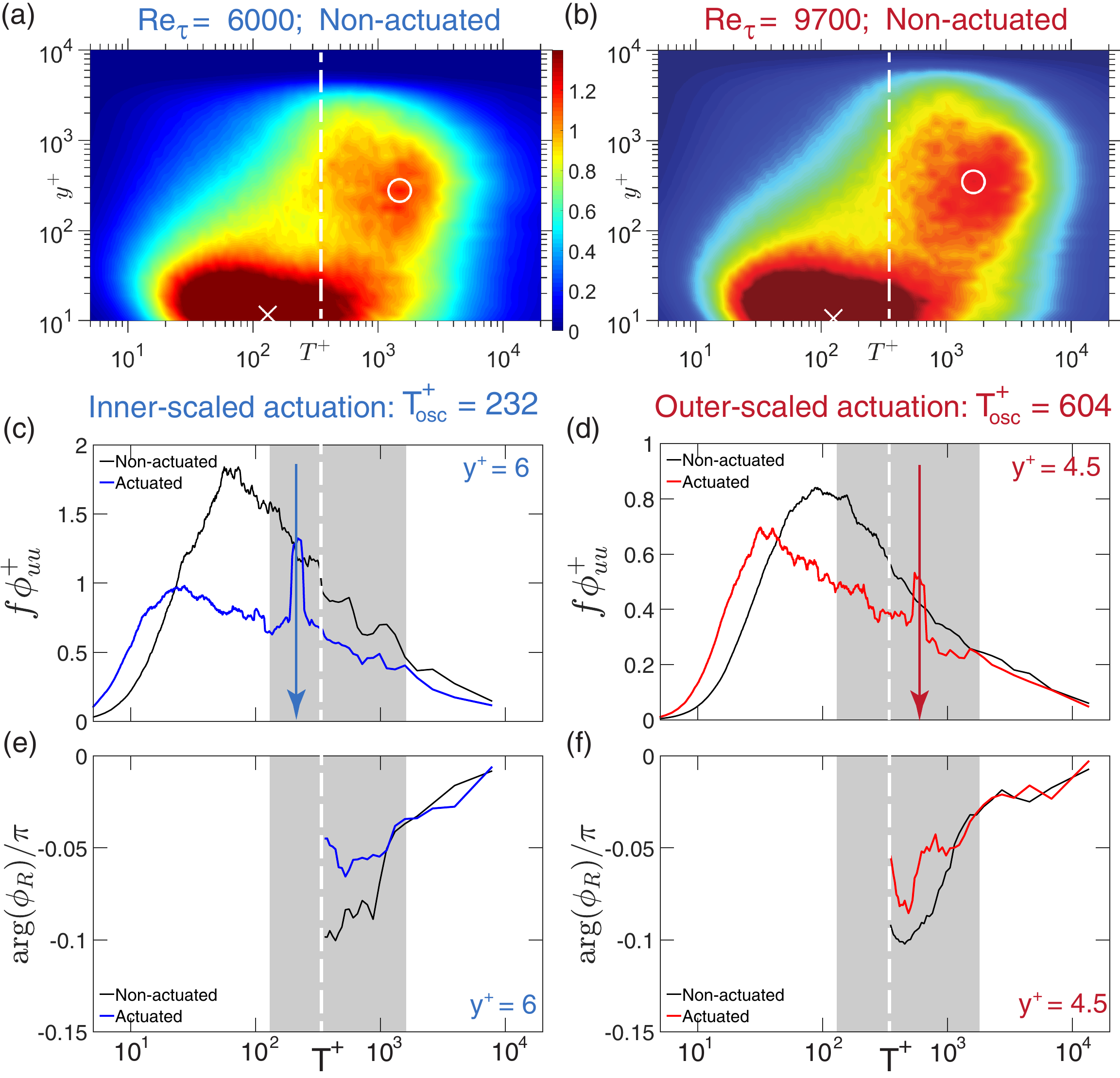}}
  \caption{(a,b) Premultiplied $u$-spectrograms, ${f}{{\phi}^{+}_{uu}}({y^{+}},{T^{+}})$ for non-actuated (canonical) turbulent boundary layers acquired by \citet{samie2018}. 
	(c,d) Premultiplied spectra of $u$(4.5$\lesssim$ $y^+$ $\lesssim$ 6) with and without wall-actuation for a specific (c) ISA and (d) OSA case.
	(e,f) Argument of the co-spectrum (${\phi}_{R}$) estimated for $T^{+}$ $>$ 350 for the same (e) ISA and (f) OSA case as in (c,d).
	Grey background represents the range of scales between the inner- (cross) and outer-peaks (circle) of the $u$-spectrograms in (a,b). 
	The arrow in (c,d) indicates $T^+_{osc}$.
    Dashed white line represents $T^{+}$ $=$ 350.}
\label{fig4}
\end{figure}

Figures \ref{fig4}(c,d) show the premultiplied $u$-spectra, ${f}{{\phi}^{+}_{uu}}$ for the non-actuated and actuated cases, where figure \ref{fig4}(c) represents the ISA strategy ($T^+_{osc}$ = 232) and figure \ref{fig4}(d) represents the OSA strategy ($T^+_{osc}$ = 604).
Here, ${{\phi}^{+}_{uu}}$ = ${{\langle}{{\widetilde{u}}(T^+)}{({{\widetilde{u}}^{*}}(T^+))}{\rangle}}/{U^{2}_{{\tau}_{o}}}$, and  `$\:{\widetilde{\;}\:}$' represents the Fourier transform in time while `${\langle}\;{\rangle}$' indicates ensemble averaging and `$^*$' indicates the complex conjugate.
The spectra for both actuated cases are attenuated across the inner ($T^+ \lesssim 350$) as well as outer ($T^+ \gtrsim 350$) scales, and  not localized to a specific scale range depending on the actuation time scale (ISA or OSA).
A broadband attenuation of the near-wall $u$-spectra, in case of the ISA strategy, has also been observed in previously published simulation data at low Reynolds numbers \citep{gatti2018,yao2019}, but the results at high Reynolds numbers are recent \citep{marusic2021}. 
At high $Re_{\tau}$, the broadband attenuation is noted across a much larger range of energetic scales, spanning the near-wall viscous(inner)-scales, the attached eddy hierarchy and the outer-scaled superstructures \citep{mathis2009,deshpande2023}, whose energetic contributions co-exist between the inner and outer peaks of the $u$-spectrogram (figures \ref{fig4}a,b).
The substantial drag reduction, which is made possible via this broadband energy attenuation, can therefore be associated with the reduced mean phase difference between the triadically coupled inner and outer scales (figures \ref{fig2},\ref{fig3}).

In addition, the maximum energy attenuation is always confined to the inner-scale region ($T^{+}$ $\lesssim$ 350) regardless of whether the spanwise oscillation targets the inner scales (figure \ref{fig4}c) or the outer scales (figure \ref{fig4}d).
For example, for OSA, the spectra at $T^{+}$ $\sim$ 100 is attenuated by $\sim$35\% on actuation, in contrast to $\sim$10-15\% attenuation at $T^+$ $\sim$ 600, which is closer to $T^{+}_{osc}$.
These observations are consistent with the discussion based on figures \ref{fig3}(e,f), where for OSA the inner-scale contributions ($u_{i}$) were found to be more severely attenuated than the outer-scale contributions. 
Hence, the manipulation of phase between the triadically coupled inner and outer scales (which enhances their coupling) appears to be correlated with the success of the OSA strategy, given it involves a substantial attenuation of the major drag contributing inner-scales despite the flow being actuated at the outer-scales (at much lower power).

While the increase in magnitude of the cross-talk term with DR in figure \ref{fig3}(d) supports the above arguments, it only gives an average estimate of the phase difference between all triadically coupled inner and outer scales.
Following \citet{jacobi2013,jacobi2017}, a more direct means to quantify the change in phase relationship on a per-scale basis is the argument ($arg$) of the co-spectrum, ${\phi}_{R}$, computed between the outer-scales ($u_{o}$) and the outer-scale `envelope' of the inner-scales (${{\epsilon}_{o}}(u_{i})$), where the co-spectrum ${\phi}_{R}$ = ${\langle}{{\widetilde{{u}_{o}}}(T^+)}\,{{{{\epsilon}_{o}}({{\widetilde{u_{i}}}^{*}}(T^+))}}{\rangle}$.  
The envelope is estimated via the Hilbert transform, with ${\phi}_{R}$ physically interpretable only for $T^{+}$ $>$ $T^{+}_{c}$, per definition (refer figures 6-7 of \citet{mathis2009} for further clarity). 
Here, ${\phi}_{R}$ is essentially a spectral equivalent of the amplitude modulation coefficient, $R({\Delta}t)$ = $\overline{ {{{u}_{o}}(t)}\, {{\epsilon}_{o}}({{u}_{i}}(t + {\Delta}t)) }$/${ { \sqrt{ \overline{ {{\epsilon}^2_{o}}(u_{i}) } } } { \sqrt{ \overline{ {{u}^2_{o}} } } } }$, which \citet{mathis2009,mathis2011} used to quantify the non-linear coupling between the inner and outer-scales.
\citet{duvvuri2015} gave an exact relationship linking $R$(${\Delta}t$ = 0) and the cross-term as: $R{\bigg(}{{\sqrt{\overline{{{\epsilon}^2_{o}}(u_{i}) } } }{ \sqrt{\overline{ {{u}^2_{o}} } } } }{\bigg)}$ = 2${\overline{ {u^2_{i}}{u_{o}} }}$, indicating that \emph{arg}(${\phi}_{R}$) would represent the scale-specific phase difference between the outer- and triadically coupled inner-scales. 

Figures \ref{fig4}(e,f) respectively show the change in scale-specific phase estimated for the same ISA and OSA cases as in figures \ref{fig4}(c,d).
Clearly, the actuation reduces the phase between $u_{o}$ and ${{\epsilon}_{o}}(u_{i})$ for both OSA and ISA cases.
This change in phase occurs nominally in the same scale-range as the attenuation of their respective premultiplied spectra (${f}{{\phi}^{+}_{uu}}$), which corresponds to the hierarchy of scales coexisting between the inner and outer peaks of the $u$-spectrogram (indicated by grey shading).
The present study, which investigates spanwise oscillations imposed at $T^+_{osc}$ across this hierarchy of scales (140 $\lesssim$ $T^+_{osc}$ $\lesssim$ 906; table \ref{tab1}), reports a significant DR that is always associated with a change in phase between the hierarchy of scales (not shown for all cases).
At very high $Re_{\tau}$ ($\gtrsim$ $\mathcal{O}$(10$^4$)), an even broader range of scales would exist between the spectral inner- and outer-peaks along with statistically significant outer-scales \citep{mathis2009,deshpande2023}. Therefore,  a significant DR could be obtained (accompanied by manipulated inter-scale phase) by actuating at any $T^+_{osc}$ within this broader energy-containing hierarchy. 
This reveals a new energy-efficient pathway that indirectly affects the major drag-producing inner-scales  through imposed actuation of the triadically coupled outer-scales.
In summary, the results from figures \ref{fig2}--\ref{fig4} are consistent in suggesting that increasing DR, for both ISA and OSA strategies, is associated with enhanced inter-scale coupling emerging through their reduced phase differences.

\section{Reynolds number variation of inter-scale phase relationships}
\label{discussion}

In the previous sections, we have argued that the success of the OSA strategy at high $Re_{\tau}$, in contrast with its failure at low $Re_{\tau}$, is associated with the energization of the outer-scales with increasing $Re_{\tau}$ \citep{marusic2021}.
This $Re_{\tau}$-dependence of the outer-scale energy is known to also enhance the non-linear coupling between the inner- and outer-scales \citep{mathis2009,mathis2011}, and consequently is bound to influence their inter-scale phase relationships (equation \ref{eq4}).
Considering the association of DR, obtained via spanwise wall oscillations, with the inter-scale phase, it is worth discussing how these relationships change in a non-actuated boundary layer for increasing $Re_{\tau}$.
To this end, we consider the hotwire dataset of \citet{marusic2015} at $Re_{\tau}$ $\sim$ 2800 and 13000, comprising streamwise velocity time-series acquired across a canonical zero-pressure gradient turbulent boundary layer.
These data were respectively acquired at the upstream and downstream ends of the long Melbourne wind tunnel test section at matched free-stream velocity (table \ref{tab1}), resulting in a significant difference in the outer-scale, $\delta$, but nominally matched viscous scale $\nu/{U_{\tau}}$ between the two cases (and hence, nominally matched hotwire spatial resolution).
Therefore, these data are well suited for investigating the non-linear `forcing'/modulation imposed by the $Re_{\tau}$-dependant outer-scales ($u_o$; $T^+$ $>$ 350), onto the `universal' (i.e. $Re_{\tau}$-invariant) inner-scales ($u_i$; $T^+$ $\lesssim$ 350).

\begin{figure}
   \captionsetup{width=1.0\linewidth}
  \centerline{\includegraphics[width=1.0\textwidth]{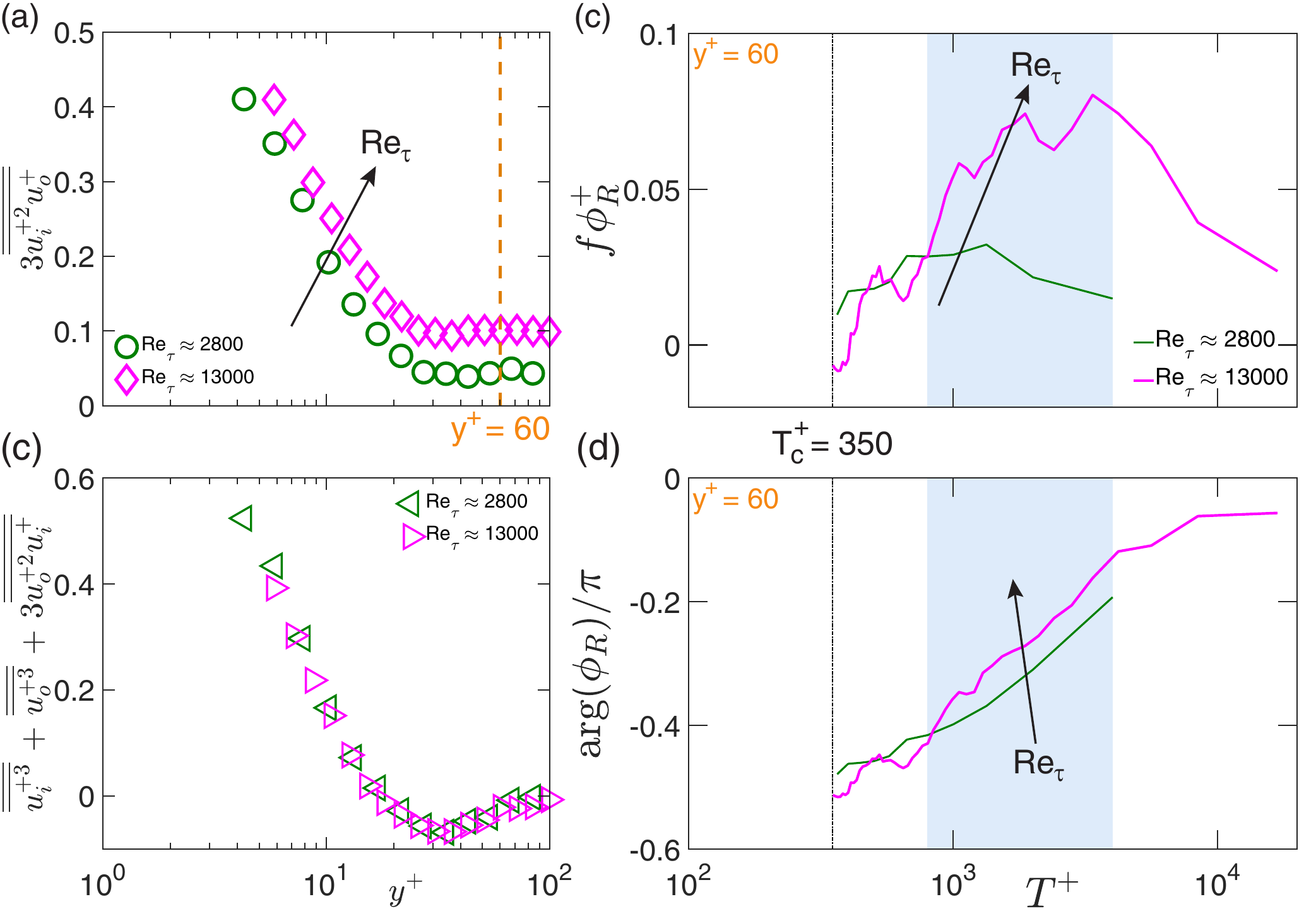}}
  \caption{(a,b) ${\mathcal{S}}_{u}$ decomposed into (a) 3${\overline{\overline{ {u^2_{i}}{u_{o}} }}}$ and (b) $\overline{\overline{u^3_{i}}}$ + $\overline{\overline{u^3_{i}}}$ + 3${\overline{\overline{ {u_{i}}{u^2_{o}} }}}$ using $T^{+}_{c}$ $=$ 350, plotted in the inner region ($y^+$ $\lesssim$ 100) using the hotwire data of \citet{marusic2015}.
	(c) Premultiplied co-spectra, ${f}{{\phi}^{+}_{R}}$ and (d) Argument of the co-spectrum, \emph{arg}(${\phi}_{R}$) estimated for $T^{+}$ $>$ 350 for canonical turbulent boundary layer data at $y^+$ $\approx$ 60.
	Light blue background represents the range of scales where the higher $Re_{\tau}$ case has a stronger inner-outer coupling than the lower $Re_{\tau}$ case. Dot-dashed black line represents $T^{+}$ $=$ 350.}
\label{fig5}
\end{figure}

Figure \ref{fig5}(a) shows the ${\mathcal{S}}_{u}$ profile decomposed into the cross-term, representative of the inner-outer coupling (3${\overline{\overline{ {u^2_{i}}{u_{o}} }}}$), while figure \ref{fig5}(b) shows the addition of the remaining terms ($\overline{\overline{u^3_{i}}}$ + $\overline{\overline{u^3_{i}}}$ + 3${\overline{\overline{ {u_{i}}{u^2_{o}} }}}$) in the inner region ($y^+$ $\lesssim$ 100).
Consistent with the previous observation of \citet{mathis2011}, 3${\overline{\overline{ {u^2_{i}}{u_{o}} }}}$ increases with $Re_{\tau}$ while all other terms are $Re_{\tau}$-invariant.
Considering the 3${\overline{\overline{ {u^2_{i}}{u_{o}} }}}$ trend with respect to (\ref{eq4}), figure \ref{fig5}(a) suggests that the increased outer-scale modulation of the inner-scales (with $Re_{\tau}$) increasingly manipulates the inter-scale phase between these scales.
This is quantified by plotting the premultiplied co-spectrum ($f{\phi}^+_{R}$) in figure \ref{fig5}(c) and \emph{arg}(${\phi}_{R}$) in figure \ref{fig5}(d) at $y^+$ $\approx$ 60, which was chosen owing to significant differences in 3${\overline{\overline{ {u^2_{i}}{u_{o}} }}}$.
$f{{\phi}^+_{R}}$ clearly indicates increased energy for $T^+$ $>$ $10^3$ for the high $Re_{\tau}$ case, and is representative of the enhanced coupling between the outer-scales and the envelope of the inner-scales.
Considering the associated \emph{arg}(${\phi}_{R}$), one can note reduced phase between $u_{o}$ and ${\epsilon_{o}}$($u_{i}$) in the same range ($T^+$ $>$ $10^3$) where the scale-specific coupling is enhanced.
The present analysis is consistent with the findings of \citet{duvvuri2015} (for a periodically forced roughness element) and that presented in $\S$\ref{results} (for imposed spanwise oscillations), thereby reaffirming that an imposed `forcing' on a turbulent boundary layer enhances the triadic coupling between inner- and outer-scales via reduction of their inter-scale phase relationship.
The fact that the inner-outer coupling increases with $Re_{\tau}$ (figure \ref{fig5}) can plausibly explain the observation of increased DR and NPS with increasing $Re_{\tau}$ for matched OSA parameters \citep{marusic2021,chandran2022}, suggesting improved efficacy of the OSA strategy at very high $Re_{\tau}$.

\section{Conclusions}
\label{future}

We find that the turbulent drag reduction (DR) achieved by spanwise wall oscillations, which provokes a broadband attenuation of drag-producing turbulent scales, is associated with an enhanced coupling between the inner and outer scales. 
This change in inter-scale coupling emerges through the manipulation of the phase relationship between these scales, wherein the actuation forces the entire range of energy-containing scales, from the inner (viscous) to the outer (inertial) scales, to be more `in-phase'.
Such a manipulation occurs regardless of whether the flow is actuated at $T^+_{osc}$ corresponding to the \emph{inner-scaled} (ISA) or the \emph{outer-scaled actuation} (OSA) pathways; both lead to a moderate DR for a high $Re_{\tau}$ flow.
The criteria to optimize this control scheme, hence, narrows down to the power required to actuate the flow, which at any given $Re_{\tau}$ is much lower in the case of OSA than ISA, making OSA the more energy-efficient pathway to high Reynolds number drag reduction \citep{marusic2021}.

The fact that manipulation of inter-scale phase is plausibly a
consequence of the imposed forcing is tested by investigating the variation in inner-outer coupling with $Re_{\tau}$, in a canonical boundary layer.
Here, the increasing non-linear modulation imposed by the $Re_{\tau}$-dependent outer-scales, on the `universal' inner-scales, is used to investigate the corresponding variation in their inter-scale phase relationships.
It is demonstrated that the increase in 3$\overline{{\overline{ {u^2_{i}}{u_{o}} }}}$, which is the sole term responsible for increasing $\mathcal{S}_u$ with $Re_{\tau}$ in a canonical flow \citep{mathis2011}, is a consequence of reduced phase between the outer-scales and outer-scale envelope of the inner-scales.
This observation is similar to that noted in a drag-reduced flow achieved via imposition of wall oscillations, where the non-linear coupling term (3$\overline{{\overline{ {u^2_{i}}{u_{o}} }}}$) is found to increase with increasing DR.
Thus, besides supporting the primary conclusion of the present study, the increase in 3$\overline{{\overline{ {u^2_{i}}{u_{o}} }}}$ with increasing $Re_{\tau}$ also suggests that the efficacy of the OSA strategy will likely improve further at higher Reynolds numbers.
These arguments encourage future work on multiple fronts, including: 
(i) investigation of the flow physics associated with enhanced inner-outer coupling at high $Re_{\tau}$, and  
(ii) testing the efficacy of practicably deployable spanwise flow oscillation schemes, such as through plasma actuators \citep{thomas2019,hehner2019,duong2021}, passive wavy walls \citep{ghebali2017}, etc., towards achieving energy-efficient drag reduction via the OSA pathway.

\section*{Appendix 1: Power calculations}
For completeness, here, we provide details of the power calculations as adapted from \citet{marusic2021}.
The net power savings (NPS) quantifies the difference between the net power cost for the non-actuated case ($P^+_0$) and the oscillating case ($P^+$ + $P^+_{in}$), and is given by: 

\begin{equation}
\label{app_eq1}
\begin{aligned}
NPS\% = 100{\times}\:{\frac{ {P^+_o} - ({P^+} + {P^+_{in}}) }{P^+_o}} = DR\%\:{-}\: \frac{P^+_{in}}{P^+_o}, 
\end{aligned}
\end{equation}
where $P^+$ and $P^+_o$ represent the power required to drive the flow over the actuated and stationary wall, respectively.
Both $P^+_o$ and $P^+$ can be estimated from the bulk velocity of the non-actuated case ($U^+_{b_o}$), i.e. $P^+_o$ = $U^+_{b_o}$ and $P^+$ = (1 - DR)$U^+_{b_o}$.

The input power ($P^+_{in}$) required to operate the spanwise oscillating mechanism is:
\begin{equation}
\label{app_eq2}
\begin{aligned}
P^+_{in} = {\frac{1}{ {t^+_{avg}}{L^+_x}{L^+_z} }} { {\int_{t^+}^{{t^+} + {t^+_{avg}}}}{\int_{0}^{{L^+_x}}}{\int_{0}^{{L^+_z}}} {w^+_{wall}} {\bigg(} {{ \frac{{\partial}w^+}{{\partial}y^+} }|}_{{y^+} = 0} {\bigg)} {d{x}^+}{d{z}^+}{d{t}^+}  }, 
\end{aligned}
\end{equation}
where $t^+_{avg}$, ${L^+_x}$ and $L^+_z$ are the averaging time, streamwise and spanwise extents of the actuated surface, respectively; $w^+_{wall}$ = ${w^+}|_{{y^+} = 0}$, ${{ {{\partial}w^+}/{{\partial}y^+} }|}_{{y^+} = 0}$ are respectively the instantaneous spanwise velocity and its wall-normal gradient at the wall. 

However, since it is not possible to measure the instantaneous spanwise velocity gradient from the present experiments, we use the laminar generalized Stokes layer (GSL) theory proposed by \citet{quadrio2011} to estimate $w^+$ following:
\begin{equation}
\label{app_eq3}
\begin{aligned}
&{w^+}({x^+},{y^+},{t^+}) =\\ 
&{A^+}{\mathcal{R}}{\bigg\{} C{e^{i({{\kappa}^+_x}{x^+} - {\omega^+_{osc}}{t^+} )} }\:\textrm{Ai}\:{\bigg[} {e^{ {\pi}i/6 }}\:({{\kappa}^+_x}[1 - DR])^{1/3} {\bigg(} {y^+} - \frac{\omega^+_{osc}}{{\kappa}^+_x[1 - DR]} - \frac{i{\kappa}^+_x}{1 - DR} {\bigg)}{\bigg]}{\bigg\}}, 
\end{aligned}
\end{equation}
where $C$ = ${\bigg\{}$\:\textrm{Ai}\:{${\bigg[}$}$i{e^{i{\pi}/3}}$\:(${\kappa}^+_x$\:[1 - DR]$)^{1/3}$\:( $\omega^+_{osc}$/${\kappa}^+_x$ + $i{\kappa^+_x}$ )/[1 $-$ DR]{${\bigg]}$}{$\bigg\}^{-1}$}, \textrm{Ai} is Airy function of the first kind, and $\mathcal{R}$ {$\{...\}$} is the real part of the argument.
\citet{marusic2021} and \citet{rouhi2022} have previously compared the $P^+_{in}$/$P^+_o$ computed using $w^+$ (from their LES data), with that estimated from the GSL theory (equation \ref{app_eq3}), and obtained good agreement. 
Notably, the maximum difference was found to be less than 1\% for their OSA cases, while it was less than 3\% for the ISA cases.
This supports usage of the GSL theory for estimating NPS for the present experiments..

\section*{Acknowledgments}
This research was funded through the Deep Science Fund of Intellectual Ventures and the Australian Research Council.
R.D. also acknowledges partial financial support by the University of Melbourne through the Melbourne Postdoctoral Fellowship.

\section*{Declaration of Interests} 

The authors report no conflict of interest.

\bibliographystyle{jfm}
\bibliography{DragReduction_InterScalePhase_bib}

\begin{thebibliography}{39}
\expandafter\ifx\csname natexlab\endcsname\relax\def\natexlab#1{#1}\fi
\def\au#1{#1} \def\ed#1{#1} \def\yr#1{#1}\def\at#1{#1}\def\jt#1{\textit{#1}}
  \def\bt#1{#1}\def\bvol#1{\textbf{#1}} \def\vol#1{#1} \def\pg#1{#1}
  \def\publ#1{#1}\def\arxiv#1{#1}\def\org#1{#1}\def\st#1{\textit{#1}}

\bibitem[Agostini {\em et~al.\/}(2014)Agostini, Touber \&
  Leschziner]{agostini2014}
{\sc \au{Agostini, L.}, \au{Touber, E.} \& \au{Leschziner, M.~A.}} \yr{2014}
  \at{Spanwise oscillatory wall motion in channel flow: drag-reduction
  mechanisms inferred from {DNS}-predicted phase-wise property variations at
  ${R}e_\tau$ $=$ $1000$}.  \jt{J. Fluid Mech.}  \bvol{743},  \pg{606--635}.

\bibitem[Akhavan {\em et~al.\/}(1993)Akhavan, Jung \&
  Mangiavacchi]{akhavan1993}
{\sc \au{Akhavan, R.}, \au{Jung, W.} \& \au{Mangiavacchi, N.}} \yr{1993}
  \at{Control of wall turbulence by high frequency spanwise oscillations}.
  \jt{AIAA paper}  \bvol{93},  \pg{3282}.

\bibitem[Auteri {\em et~al.\/}(2010)Auteri, Baron, Belan, Campanardi \&
  Quadrio]{auteri2010}
{\sc \au{Auteri, F.}, \au{Baron, A.}, \au{Belan, M.}, \au{Campanardi, G.} \&
  \au{Quadrio, M.}} \yr{2010}  \at{Experimental assessment of drag reduction by
  traveling waves in a turbulent pipe flow}.  \jt{Phys. Fluids}
  \bvol{22}~(11),  \pg{115103}.

\bibitem[Baron \& Quadrio(1995)]{baron1995}
{\sc \au{Baron, A.} \& \au{Quadrio, M.}} \yr{1995}  \at{Turbulent drag
  reduction by spanwise wall oscillations}.  \jt{Applied Scientific Research}
  \bvol{55},  \pg{311--326}.

\bibitem[Chandran {\em et~al.\/}(2023)Chandran, Zampiron, Rouhi, Fu, Wine,
  Holloway, Smits \& Marusic]{chandran2022}
{\sc \au{Chandran, D.}, \au{Zampiron, A.}, \au{Rouhi, A.}, \au{Fu, M.~K.},
  \au{Wine, D.}, \au{Holloway, B.}, \au{Smits, A.~J.} \& \au{Marusic, I.}}
  \yr{2023}  \at{Turbulent drag reduction by spanwise wall forcing. {P}art 2.
  {H}igh-{R}eynolds-number experiments}.  \jt{Journal of Fluid Mechanics}
  \bvol{968},  \pg{A7}.

\bibitem[Choi {\em et~al.\/}(1998)Choi, DeBisschop \& Clayton]{choi1998}
{\sc \au{Choi, K.S.}, \au{DeBisschop, J.R.} \& \au{Clayton, B.R.}} \yr{1998}
  \at{Turbulent boundary-layer control by means of spanwise-wall oscillation}.
  \jt{AIAA journal}  \bvol{36}~(7),  \pg{1157--1163}.

\bibitem[Deck {\em et~al.\/}(2014)Deck, Renard, Laraufie \& Weiss]{deck2014}
{\sc \au{Deck, S.}, \au{Renard, N.}, \au{Laraufie, R.} \& \au{Weiss, P.}}
  \yr{2014}  \at{Large-scale contribution to mean wall shear stress in
  high-{R}eynolds-number flat-plate boundary layers up to 13650}.  \jt{J. Fluid
  Mech.}  \bvol{743},  \pg{202--248}.

\bibitem[Deshpande {\em et~al.\/}(2023)Deshpande, de~Silva \&
  Marusic]{deshpande2023}
{\sc \au{Deshpande, R.}, \au{de~Silva, C.~M.} \& \au{Marusic, I.}} \yr{2023}
  \at{Evidence that superstructures comprise self-similar coherent motions in
  high reynolds number boundary layers}.  \jt{Journal of Fluid Mechanics}
  \bvol{969},  \pg{A10}.

\bibitem[Duong {\em et~al.\/}(2021)Duong, Corke \& Thomas]{duong2021}
{\sc \au{Duong, A.H.}, \au{Corke, T.C.} \& \au{Thomas, F.O.}} \yr{2021}
  \at{Characteristics of drag-reduced turbulent boundary layers with
  pulsed-direct-current plasma actuation}.  \jt{Journal of Fluid Mechanics}
  \bvol{915},  \pg{A113}.

\bibitem[Duvvuri \& McKeon(2017)]{duvvuri2017}
{\sc \au{Duvvuri, S.} \& \au{McKeon, B.}} \yr{2017}  \at{Phase relations in a
  forced turbulent boundary layer: implications for modelling of high
  {R}eynolds number wall turbulence}.  \jt{Phil. Trans. R. Soc. A:}
  \bvol{375}~(2089),  \pg{20160080}.

\bibitem[Duvvuri \& McKeon(2015)]{duvvuri2015}
{\sc \au{Duvvuri, S.} \& \au{McKeon, B.~J.}} \yr{2015}  \at{Triadic scale
  interactions in a turbulent boundary layer}.  \jt{J. Fluid Mech.}
  \bvol{767},  \pg{{R}4}.

\bibitem[Gatti \& Quadrio(2016)]{gatti2016}
{\sc \au{Gatti, D.} \& \au{Quadrio, M.}} \yr{2016}  \at{Reynolds-number
  dependence of turbulent skin-friction drag reduction induced by spanwise
  forcing}.  \jt{J. Fluid Mech.}  \bvol{802},  \pg{553--582}.

\bibitem[Gatti {\em et~al.\/}(2018)Gatti, Stroh, Frohnapfel \&
  Hasegawa]{gatti2018}
{\sc \au{Gatti, D.}, \au{Stroh, A.}, \au{Frohnapfel, B.} \& \au{Hasegawa, Y.}}
  \yr{2018}  \at{Predicting turbulent spectra in drag-reduced flows}.  \jt{Flow
  Turbulence Combust.}  \bvol{100}~(4),  \pg{1081--1099}.

\bibitem[Ghebali {\em et~al.\/}(2017)Ghebali, Chernyshenko \&
  Leschziner]{ghebali2017}
{\sc \au{Ghebali, S.}, \au{Chernyshenko, S.~I.} \& \au{Leschziner, M.~A.}}
  \yr{2017}  \at{Can large-scale oblique undulations on a solid wall reduce the
  turbulent drag?}  \jt{Physics of Fluids}  \bvol{29}~(10),  \pg{105102}.

\bibitem[Hamilton {\em et~al.\/}(1995)Hamilton, Kim \& Waleffe]{hamilton1995}
{\sc \au{Hamilton, J.~M.}, \au{Kim, J.} \& \au{Waleffe, F.}} \yr{1995}
  \at{Regeneration mechanisms of near-wall turbulence structures}.  \jt{J.
  Fluid Mech.}  \bvol{287},  \pg{317--348}.

\bibitem[Hehner {\em et~al.\/}(2019)Hehner, Gatti \& Kriegseis]{hehner2019}
{\sc \au{Hehner, M.T.}, \au{Gatti, D.} \& \au{Kriegseis, J.}} \yr{2019}
  \at{Stokes-layer formation under absence of moving parts—a novel
  oscillatory plasma actuator design for turbulent drag reduction}.
  \jt{Physics of Fluids}  \bvol{31}~(5),  \pg{051701}.

\bibitem[Jacobi \& McKeon(2013)]{jacobi2013}
{\sc \au{Jacobi, I.} \& \au{McKeon, B.~J.}} \yr{2013}  \at{Phase relationships
  between large and small scales in the turbulent boundary layer}.  \jt{Exp.
  Fluids}  \bvol{54}~(3),  \pg{1--13}.

\bibitem[Jacobi \& McKeon(2017)]{jacobi2017}
{\sc \au{Jacobi, I.} \& \au{McKeon, B.~J.}} \yr{2017}  \at{Phase-relationships
  between scales in the perturbed turbulent boundary layer}.  \jt{J.
  Turbulence}  \bvol{18}~(12),  \pg{1120--1143}.

\bibitem[Jim{\'e}nez \& Moin(1991)]{jimenez1991}
{\sc \au{Jim{\'e}nez, J.} \& \au{Moin, P.}} \yr{1991}  \at{The minimal flow
  unit in near-wall turbulence}.  \jt{J. Fluid Mech.}  \bvol{225},
  \pg{213--240}.

\bibitem[Jim{\'e}nez \& Pinelli(1999)]{jimenez1999}
{\sc \au{Jim{\'e}nez, J.} \& \au{Pinelli, A.}} \yr{1999}  \at{The autonomous
  cycle of near-wall turbulence}.  \jt{J. Fluid Mech.}  \bvol{389},
  \pg{335--359}.

\bibitem[Karniadakis \& Choi(2003)]{karniadakis2003}
{\sc \au{Karniadakis, G.E.} \& \au{Choi, K.S.}} \yr{2003}  \at{Mechanisms on
  transverse motions in turbulent wall flows}.  \jt{Ann. Rev. Fluid Mech.}
  \bvol{35}~(1),  \pg{45--62}.

\bibitem[Kim(2011)]{kim2011}
{\sc \au{Kim, J.}} \yr{2011}  \at{Physics and control of wall turbulence for
  drag reduction}.  \jt{Philosophical Transactions of the Royal Society A:
  Mathematical, Physical and Engineering Sciences}  \bvol{369}~(1940),
  \pg{1396--1411}.

\bibitem[Marusic {\em et~al.\/}(2021)Marusic, Chandran, Rouhi, Fu, Wine,
  Holloway, Chung \& Smits]{marusic2021}
{\sc \au{Marusic, I.}, \au{Chandran, D.}, \au{Rouhi, A.}, \au{Fu, M.~K.},
  \au{Wine, D.}, \au{Holloway, B.}, \au{Chung, D.} \& \au{Smits, A.~J.}}
  \yr{2021}  \at{An energy-efficient pathway to turbulent drag reduction}.
  \jt{Nat. Commun.}  \bvol{12}~(1),  \pg{1--8}.

\bibitem[Marusic {\em et~al.\/}(2015)Marusic, Chauhan, Kulandaivelu \&
  Hutchins]{marusic2015}
{\sc \au{Marusic, I.}, \au{Chauhan, K.A.}, \au{Kulandaivelu, V.} \&
  \au{Hutchins, N.}} \yr{2015}  \at{Evolution of zero-pressure-gradient
  boundary layers from different tripping conditions}.  \jt{J. Fluid Mech.}
  \bvol{783},  \pg{379--411}.

\bibitem[Mathis {\em et~al.\/}(2009)Mathis, Hutchins \& Marusic]{mathis2009}
{\sc \au{Mathis, R.}, \au{Hutchins, N.} \& \au{Marusic, I.}} \yr{2009}
  \at{Large-scale amplitude modulation of the small-scale structures in
  turbulent boundary layers}.  \jt{J. Fluid Mech.}  \bvol{628},  \pg{311--337}.

\bibitem[Mathis {\em et~al.\/}(2011)Mathis, Marusic, Hutchins \&
  Sreenivasan]{mathis2011}
{\sc \au{Mathis, R.}, \au{Marusic, I.}, \au{Hutchins, N.} \& \au{Sreenivasan,
  K.~R.}} \yr{2011}  \at{The relationship between the velocity skewness and the
  amplitude modulation of the small scale by the large scale in turbulent
  boundary layers}.  \jt{Phys. Fluids}  \bvol{23}~(12),  \pg{121702}.

\bibitem[Pal {\em et~al.\/}(1989)Pal, Deutsch \& Merkle]{pal1989}
{\sc \au{Pal, S.}, \au{Deutsch, S.} \& \au{Merkle, C.L.}} \yr{1989}  \at{A
  comparison of shear stress fluctuation statistics between microbubble
  modified and polymer modified turbulent boundary layers}.  \jt{Physics of
  Fluids A: Fluid Dynamics}  \bvol{1}~(8),  \pg{1360--1362}.

\bibitem[Quadrio \& Ricco(2011)]{quadrio2011}
{\sc \au{Quadrio, M.} \& \au{Ricco, P.}} \yr{2011}  \at{The laminar generalized
  {S}tokes layer and turbulent drag reduction}.  \jt{J. Fluid Mech.}
  \bvol{667},  \pg{135--157}.

\bibitem[Quadrio {\em et~al.\/}(2009)Quadrio, Ricco \& Viotti]{quadrio2009}
{\sc \au{Quadrio, M.}, \au{Ricco, P.} \& \au{Viotti, C.}} \yr{2009}
  \at{Streamwise-travelling waves of spanwise wall velocity for turbulent drag
  reduction}.  \jt{J. Fluid Mech.}  \bvol{627},  \pg{161--178}.

\bibitem[Ricco {\em et~al.\/}(2021)Ricco, Skote \& Leschziner]{ricco2021}
{\sc \au{Ricco, P.}, \au{Skote, M.} \& \au{Leschziner, M.~A.}} \yr{2021}  \at{A
  review of turbulent skin-friction drag reduction by near-wall transverse
  forcing}.  \jt{Prog. Aero. Sci.}  \bvol{123},  \pg{100713}.

\bibitem[Ricco \& Wu(2004)]{ricco2004}
{\sc \au{Ricco, P.} \& \au{Wu, S.}} \yr{2004}  \at{On the effects of lateral
  wall oscillations on a turbulent boundary layer}.  \jt{Exp. Thermal Fluid
  Sci.}  \bvol{29}~(1),  \pg{41--52}.

\bibitem[Rouhi {\em et~al.\/}(2023)Rouhi, Fu, Chandran, Zampiron, Smits \&
  Marusic]{rouhi2022}
{\sc \au{Rouhi, A.}, \au{Fu, M.~K.}, \au{Chandran, D.}, \au{Zampiron, A.},
  \au{Smits, A.~J.} \& \au{Marusic, I.}} \yr{2023}  \at{Turbulent drag
  reduction by spanwise wall forcing. {P}art 1. {L}arge-eddy simulations}.
  \jt{Journal of Fluid Mechanics}  \bvol{968},  \pg{A6}.

\bibitem[Samie {\em et~al.\/}(2018)Samie, Marusic, Hutchins, Fu, Fan, Hultmark
  \& Smits]{samie2018}
{\sc \au{Samie, M.}, \au{Marusic, I.}, \au{Hutchins, N.}, \au{Fu, M.~K.},
  \au{Fan, Y.}, \au{Hultmark, M.} \& \au{Smits, A.~J.}} \yr{2018}  \at{Fully
  resolved measurements of turbulent boundary layer flows up to ${R}e_{\tau}$ =
  $20$ $000$}.  \jt{J. Fluid Mech.}  \bvol{851},  \pg{391--415}.

\bibitem[Schoppa \& Hussain(2002)]{schoppa2002}
{\sc \au{Schoppa, W.} \& \au{Hussain, F.}} \yr{2002}  \at{Coherent structure
  generation in near-wall turbulence}.  \jt{J. Fluid Mech.}  \bvol{453},
  \pg{57--108}.

\bibitem[Skote {\em et~al.\/}(2019)Skote, Mishra \& Wu]{skote2019wall}
{\sc \au{Skote, M.}, \au{Mishra, M.} \& \au{Wu, Y.}} \yr{2019}  \at{Wall
  oscillation induced drag reduction zone in a turbulent boundary layer}.
  \jt{Flow Turbul. Comb.}  \bvol{102}~(3),  \pg{641--666}.

\bibitem[Talluru(2013)]{talluru2013}
{\sc \au{Talluru, Krishna}} \yr{2013}  \at{Manipulating large-scale structures
  in a turbulent boundary layer using a wall-normal jet}. PhD thesis,
  University of Melbourne, Department of Mechanical Engineering.

\bibitem[Thomas {\em et~al.\/}(2019)Thomas, Corke, Duong, Midya \&
  Yates]{thomas2019}
{\sc \au{Thomas, F.O.}, \au{Corke, T.C.}, \au{Duong, A.}, \au{Midya, S.} \&
  \au{Yates, K.}} \yr{2019}  \at{Turbulent drag reduction using pulsed-dc
  plasma actuation}.  \jt{Journal of Physics D: Applied Physics}
  \bvol{52}~(43),  \pg{434001}.

\bibitem[Waleffe {\em et~al.\/}(1993)Waleffe, Kim \& Hamilton]{waleffe1993}
{\sc \au{Waleffe, F}, \au{Kim, J.} \& \au{Hamilton, J.~M.}} \yr{1993} On the
  origin of streaks in turbulent shear flows.  \bt{In {\em Turbulent Shear
  Flows 8: Selected Papers from the Eighth International Symposium on Turbulent
  Shear Flows, Munich, Germany, September 9--11, 1991\/}},  \pg{pp. 37--49}.
  Springer.

\bibitem[Yao {\em et~al.\/}(2019)Yao, Chen \& Hussain]{yao2019}
{\sc \au{Yao, J.}, \au{Chen, X.} \& \au{Hussain, F.}} \yr{2019}  \at{Reynolds
  number effect on drag control via spanwise wall oscillation in turbulent
  channel flows}.  \jt{Physics of Fluids}  \bvol{31}~(8),  \pg{085108}.

\end{thebibliography}

\end{document}